\documentclass[10pt, conference, letterpaper]{IEEEtran}
\IEEEoverridecommandlockouts
\usepackage{cite}
\usepackage{amsmath,amssymb,amsfonts}
\usepackage{algorithmic}
\usepackage{graphicx}
\usepackage{textcomp}
\usepackage{xcolor}
\usepackage{svg}
\usepackage{braket}
\usepackage{booktabs}
\usepackage{caption}

\usepackage{multirow}
\usepackage{adjustbox}
\def\BibTeX{{\rm B\kern-.05em{\sc i\kern-.025em b}\kern-.08em
    T\kern-.1667em\lower.7ex\hbox{E}\kern-.125emX}}

\usepackage{graphicx}
\graphicspath{ {./fig/} }
\usepackage{multirow}
\usepackage{booktabs}

\usepackage{cite}
\usepackage{amsfonts}
\usepackage{amssymb}   % For more math symbols
\usepackage{amsmath}
\usepackage{algorithmic}
\usepackage{adjustbox}
\usepackage{caption}
\usepackage{wrapfig}
\usepackage{enumitem}

\usepackage{xcolor}
\usepackage[ruled,vlined]{algorithm2e}

\SetCommentSty{mycommfont}
\usepackage{soul}
\usepackage{mathtools}

\usepackage{colortbl}

\usepackage{makecell}

\usepackage{textcomp}

\begin{document}
\abovedisplayskip=5pt
\abovedisplayshortskip=5pt
\belowdisplayskip=5pt
\belowdisplayshortskip=5pt

\title{Input-Aware Dynamic Backdoor Attack Against Quantum Neural Networks}

\author{
Junrui Zhang$^{1}$, Zemin Chen$^{1}$, Lusi Li$^{1}$, Mohammad Ghasemigol$^{2}$, Daniel Takabi$^{2}$ and Rui Ning$^{1}$ \\[0.6em]
$^{1}$Department of Computer Science, Old Dominion University, Norfolk, VA, USA \\
$^{2}$School of Cybersecurity , Old Dominion University, Norfolk, VA, USA
}

% \author[1]{Junrui Zhang}
% \author[1]{Zemin Chen}
% \author[1]{Steven Chen}
% \author[3]{Liuwan Zhu}
% \author[1]{Lusi Li}
% \author[1]{Rui Ning}
% \author[5]{Hongyi Wu}
% \author[2]{Chunsheng Xin}
% \author[1]{Rui Ning}

% \affil[1]{Department of CS, Old Dominion University, Norfolk, VA 23529, USA}
% \affil[2]{Department of ECE, Old Dominion University, Norfolk, VA 23529, USA}
% \affil[3]{Department of ECE, University of Hawaii at Manoa, Honolulu, HI 96822, USA}
% \affil[5]{Department of ECE, University of Arizona, Tucson, AZ 85721, USA}

\maketitle
\raggedbottom

\begin{abstract}
Quantum Neural Networks (QNNs) are emerging as a promising framework for quantum machine learning on near-term quantum devices, but their security risks remain insufficiently understood. Recent studies have shown that QNNs are vulnerable to backdoor attacks, yet existing quantum backdoors mostly rely on a fixed trigger shared by all poisoned inputs. This fixed-trigger design is a major weakness because many backdoor defenses are built to detect or weaken the repeated patterns such triggers leave in data representations. Although input-aware dynamic backdoors have been studied in classical neural networks to overcome this limitation, directly transferring them to QNNs is difficult because quantum learning introduces new obstacles. In particular, measurement compresses the post-ansatz quantum state into a limited classical output, which weakens the supervision signal for training a trigger generator, while individual density matrices fluctuate strongly with the input and make per-sample contrastive learning unstable. To address these challenges, we propose Q-DIBA, the first input-aware dynamic backdoor attack designed specifically for QNNs. Q-DIBA jointly trains a classical trigger generator and a victim QNN through a three-mode mini-batch strategy that supports clean behavior, attack activation, and trigger specificity. To provide stable quantum-level supervision, Q-DIBA introduces an ensemble density contrastive loss that operates on post-ansatz quantum states before measurement and contrasts mode-averaged density matrices rather than individual samples. Experiments on MNIST and F-MNIST across multiple QNN architectures show that Q-DIBA achieves high clean accuracy, strong attack success, and high cross-trigger accuracy, demonstrating effectiveness, stealthiness, and input specificity. The attack also remains resilient against representative defenses, including visual inspection, spectral-signature detection, and fine-tuning, suggesting that input-aware quantum backdoors represent a practical and important threat to secure QNN deployment.

% \noindent  Quantum Neural Networks (QNNs) has emerged as a promising paradigm for quantum machine learning in the Noisy Intermediate-Scale Quantum (NISQ) era. However, QNNs are vulnerable to security threats, particularly backdoor attacks. Existing quantum backdoor attacks rely on a single fixed trigger pattern shared across poisoned inputs, making them easy to detect and mitigate with current defense methods. Directly adapting input-aware backdoors to the quantum setting is non-trivial: post-measurement supervision provides only a contracted learning signal, and per-sample density matrices carry strong input-driven fluctuations that overwhelm training objectives. To tackle this, we propose Q-DIBA, a dynamic input-aware backdoor attack against QNNs that addresses both obstacles. Q-DIBA jointly trains a classical trigger generator and a victim QNN under a three-mode mini-batch strategy. The training is supervised by an ensemble density contrastive loss that operates on post-ansatz quantum states and contrasts ensemble-averaged density matrices, recovering a stable gradient for the joint optimization. Experiments on MNIST and Fashion-MNIST show that Q-DIBA outperforms other baselines to achieve high clean accuracy (CA), attack success rate (ASR) and cross-trigger accuracy (CTA) across multiple QNN architectures, demonstrating its stealthiness, effectiveness, and specificity. The failure of three typical defense to defeat Q-DIBA further demonstrates its resilience against existing backdoor defenses.
\end{abstract}


\vspace{-2mm}\section{Introduction}\vspace{-2mm}
Machine learning has driven dramatic progress and been broadly adopted in real-world applications such as computer vision, natural language processing, and autonomous systems. Building on this momentum, quantum computing has emerged as a paradigm with the potential to fundamentally enhance machine learning capabilities. By exploiting quantum phenomena such as superposition and entanglement, quantum systems embed classical data into exponentially large Hilbert spaces, yielding feature representations that capture patterns beyond the reach of classical methods~\cite{havlivcek2019supervised}.

In the current noisy intermediate-scale quantum (NISQ) era~\cite{preskill2018quantum}, quantum devices are limited by noise and shallow circuit depths. To efficiently exploit quantum advantages within NISQ hardware constraints, Quantum Neural Networks (QNNs)~\cite{biamonte2017quantum} have been proposed as a promising framework that employs parameterized quantum circuits (PQCs) trained by classical optimizers. By creating a hybrid quantum-classical pipeline, QNNs are capable of learning complex data patterns while remaining executable on near-term quantum hardware.

While QNNs inherit the power of classical algorithms, they also inherit their critical security weaknesses, most notably susceptibility to backdoor attacks~\cite{gu2017badnets,chen2017targeted,nguyen2020input}. By injecting hidden trigger patterns into the training data or model parameters, an attacker can poison the model to misbehave whenever the trigger is present during inference. The poisoned model still behave normally on benign inputs, which makes such attacks particularly stealthy and difficult to detect through standard validation procedures. This vulnerability poses significant risks for the deployment of QNNs in security-critical applications.

\vspace{1mm}
\noindent\textbf{The limitation of fixed-trigger quantum backdoors.}
Recent studies have begun to explore backdoor attacks in QNNs. Circuit-level attacks such as QTrojan and QDoor insert malicious behavior into the quantum circuit and usually require white-box access to the circuit architecture~\cite{chu2023qtrojan,chu2023qdoor}. Data-poisoning attacks such as Q-FGSM, QUAP and HarmQ craft poisoned training samples or adversarial perturbations under weaker access assumptions~\cite{huang2023backdoor,zhao2025black,zhang2026harmq}. Although these methods differ in mechanism and threat model, they share one key design choice: \textit{the same trigger is applied to every poisoned input.}
This fixed-trigger design limits the strength of the attack. It creates a repeated pattern in the data space or representation space, and this regularity is exactly what many backdoor defenses exploit. Neural Cleanse searches for a small trigger that consistently induces one target class~\cite{wang2019neural}. Spectral Signature detects poisoned samples by finding abnormal feature directions~\cite{tran2018spectral}. Fine-tuning can also weaken a fixed-trigger backdoor because the malicious behavior is tied to one reusable pattern. Therefore, a stronger quantum backdoor should not rely on a universal trigger.

% Recent work has investigated how backdoor attacks manifest specifically in the quantum setting. Circuit-level attacks such as QTrojan~\cite{chu2023qtrojan} and QDoor~\cite{chu2023qdoor} embed malicious behavior directly into the quantum architecture, leveraging white-box access to the circuit topology. Data-poisoning attacks, including Q-FGSM~\cite{huang2023backdoor}, QUAP~\cite{zhao2025black} and HarmQ~\cite{zhang2026harmq}, instead craft trigger patterns through adversarial perturbations and inject them into the training set. Despite their differences in mechanism and threat assumptions, the existing quantum backdoor attacks share a single underlying design choice: a fixed, uniform trigger pattern is applied identically to every poisoned input. Classical backdoor research has shown that this uniformity is the structural property exploited by most existing defenses~\cite{wang2019neural,tran2018spectral}: Neural Cleanse and spectral signatures search for the regularity that a single trigger introduces into the feature space, and fine-tuning gradually erodes a fixed pattern because there is only one pattern to erode. An attack that breaks the fixed-trigger structure would therefore invalidate the assumptions underlying these defenses.

\vspace{1mm}
\noindent\textbf{Why input-aware backdoors matter for QNNs.}
Classical backdoor research has already recognized the weakness of fixed triggers and developed input-aware dynamic backdoor attacks~\cite{nguyen2020input}. In these attacks, a trigger generator produces a different trigger for each input. The trigger is not a reusable patch shared by all poisoned samples. Instead, it is bound to its source input, so a trigger generated from one sample should not activate the backdoor on another sample. Later studies further extend this idea through flexible generators and sample-specific transformations~\cite{salem2020dynamic,nguyen2021wanet}.

The existence of input-aware backdoors in classical neural networks does not settle whether the same threat exists for QNNs. This raises an important security question: are QNNs also vulnerable to input-aware backdoors, or does the quantum learning pipeline prevent them? This gap matters for two reasons. First, fixed-trigger quantum attacks may underestimate the capability of a realistic training-stage attacker. Second, defenses that appear effective against fixed quantum triggers may fail once the trigger becomes sample-specific. We therefore study input-aware backdoors in QNNs to close this missing threat model and to examine whether QNN security relies on assumptions that are already known to be fragile in classical learning.

% To overcome these limitations, the classical backdoor community has shifted toward input-aware dynamic backdoor attacks~\cite{nguyen2020input}, in which a learned generator emits a distinct, sample-specific trigger for every input. Such attacks invalidate the assumption underlying fixed-trigger defenses and substantially raise the difficulty of detection and mitigation. Nguyen and Tran~\cite{nguyen2020input} train a trigger generator with a diversity loss and a cross-trigger test that enforces non-reusability, so each generated pattern activates the backdoor only on its source input. Salem et al.~\cite{salem2022dynamic} further generalize this idea through GAN-style generators that produce triggers with random or label-conditioned patterns and locations. WaNet~\cite{nguyen2021wanet} replaces additive noise triggers with imperceptible warping fields, demonstrating that sample-specific deformations can also evade visual inspection and the leading machine level defenses. However, while these works establish input-aware backdoors as a mature attack family in the classical setting, input-aware dynamic backdoor mechanisms remain largely unexplored in the quantum domain.

\vspace{1mm}
\noindent\textbf{Why classical dynamic attacks cannot be directly transferred.}
Although the motivation comes from classical input-aware backdoors, the attack method cannot be directly copied into the quantum setting. Classical dynamic backdoor attacks usually train a trigger generator through image-space objectives and output-level supervision. A QNN introduces a different information flow. After the PQCs transforms the encoded input, measurement maps the post-ansatz quantum state into a small set of classical values. This quantum-to-classical readout can discard coherence and multi-qubit correlations that are still present in the quantum state. As a result, a loss computed only on the measured output provides an attenuated supervision signal for training the trigger generator.

A natural way to recover this lost signal is to compute the auxiliary loss on the post-ansatz quantum state itself, before the measurement layer discards it. However, this introduces a second obstacle. A single density matrix $\rho_i = |\psi_i\rangle\langle\psi_i|$ can vary strongly with the input identity. If contrastive learning is applied to individual samples, this input-dependent fluctuation can dominate the structural signal needed for backdoor learning. The resulting gradients become unstable and cannot reliably guide the joint training of the trigger generator and the victim QNN. These two obstacles show that input-aware backdoors in QNNs require a quantum-native design rather than a direct adaptation of classical trigger generators.

% Bringing input-aware backdoor attacks into the quantum setting, however, is not a matter of straightforward adaptation. Two limitations of the quantum learning pipeline obstruct any direct implementation of the classical approach: 1) First, supervision at the QNNs' logits provides only a weakened learning signal. The path from the post-ansatz quantum state to the classifier output is a sequence of completely positive trace preserving (CPTP) operations, that are weakly contractive under the trace distance. And in our pipeline measurement further discards all off-diagonal coherence and most multi-qubit correlations carried by the state. The loss computed on the classifier output therefore operates on a substantially attenuated class-conditional signal, leaving the trigger generator with too weak a gradient. 2) Second, per-sample quantum observations fluctuate strongly: any single $\rho_i = |\psi_i\rangle\langle\psi_i|$ carries substantial fluctuation that obscures the structural information a contrastive objective relies on. A per-sample contrastive loss therefore operates at a low signal-to-noise ratio and yields gradients too unstable to drive consistent learning.

\vspace{1mm}
\noindent\textbf{Our approach.}
To address this gap, we propose Q-DIBA, the first input-aware dynamic backdoor attack designed specifically for QNNs. Q-DIBA jointly trains a classical trigger generator and a victim QNN. The generator produces a sample-specific trigger, while the victim QNN learns three behaviors through a three-mode mini-batch construction. Clean samples preserve normal model utility. Backdoor samples teach the model to predict the attacker target when an input carries its own trigger. Cross-trigger samples carry a trigger generated from a different input but keep their own ground-truth label, so the model must predict correctly despite the trigger, which forces the trigger to be input-specific rather than universal.

% To address these limitations, we propose Q-DIBA, the first input-aware dynamic backdoor attack designed specifically for QNNs. Q-DIBA moves the contrastive supervision upstream of the measurement, defining the loss directly on the post-ansatz quantum state, and contrasts ensemble-averaged density matrices rather than individual samples to cancel per-sample fluctuations and recover the systematic signal.

\noindent\textbf{This paper makes the following key contributions:}
\begin{enumerate}[leftmargin=*]
    \item We identify input-aware dynamic backdoor attacks as a missing threat model for QNNs. This threat is important because existing quantum backdoors still rely on fixed triggers, while fixed triggers are the key assumption used by many defenses.
    \item We analyze and identify two quantum-specific obstacles that prevent the direct adaptation of classical input-aware backdoor attacks to QNNs: measurement-induced signal contraction and per-sample observation fluctuation.
    \item We propose Q-DIBA, the first input-aware dynamic backdoor attack against QNNs, which addresses both obstacles through pre-measurement contrastive supervision on ensemble-averaged density matrices.
    \item We empirically validate Q-DIBA on MNIST and F-MNIST across multiple QNN architectures, achieving high clean accuracy, attack success rate, and cross-trigger accuracy that outperform existing backdoor attacks, while maintaining strong robustness against existing backdoor defenses.
\end{enumerate}

The rest of the paper is organized as follows: Section II reviews QNNs preliminaries and backdoor background. Section III presents the threat model and the framework design. Section IV details experimental settings and results, including robustness, comparisons, and ablations. Section V concludes.

\vspace{-1mm}\section{Preliminaries}\vspace{-1mm}
\label{sec:back}

\subsection{Quantum Bits and Quantum States}

The qubit is the elementary information unit of a quantum system, generalizing the classical bit by allowing arbitrary superpositions of its two computational basis states. A single-qubit pure state is written as
\begin{equation}
\ket{\psi} = \alpha \ket{0} + \beta \ket{1},
\quad |\alpha|^2 + |\beta|^2 = 1,
\end{equation}
where the complex amplitudes $\alpha, \beta$ encode both the relative weighting and the phase of each basis component, and the normalization condition guarantees that $|\alpha|^2$ and $|\beta|^2$ form a valid probability distribution over the two possible measurement outcomes. Due to this characteristics, a qubit can carry a continuum of states across the entire $\alpha,\beta$ space, allowing it to represent richer information in the same physical resource. A measurement in the computational basis collapses $\ket{\psi}$ to $\ket{0}$ with probability $|\alpha|^2$ or to $\ket{1}$ with probability $|\beta|^2$, irreversibly destroying the superposition.

Two qubits become entangled when their joint state cannot be written as a tensor product of individual qubit states. A canonical example is the Bell state $\ket{\Phi^+}_{AB}=\frac{1}{\sqrt{2}}\left(\ket{0}_A\ket{0}_B+\ket{1}_A\ket{1}_B\right)$, in which measuring one qubit instantaneously fixes the outcome of the other. Superposition and entanglement together give $n$-qubit systems access to a $2^n$-dimensional Hilbert space, and this exponential representational capacity is the resource that quantum machine learning exploits: data embedded into a quantum state can be processed in a feature space far larger than what a classical model of comparable size can access.

\subsection{Quantum Neural Networks}
\begin{figure}[h!]
    \centering
    \includegraphics[width=\linewidth]{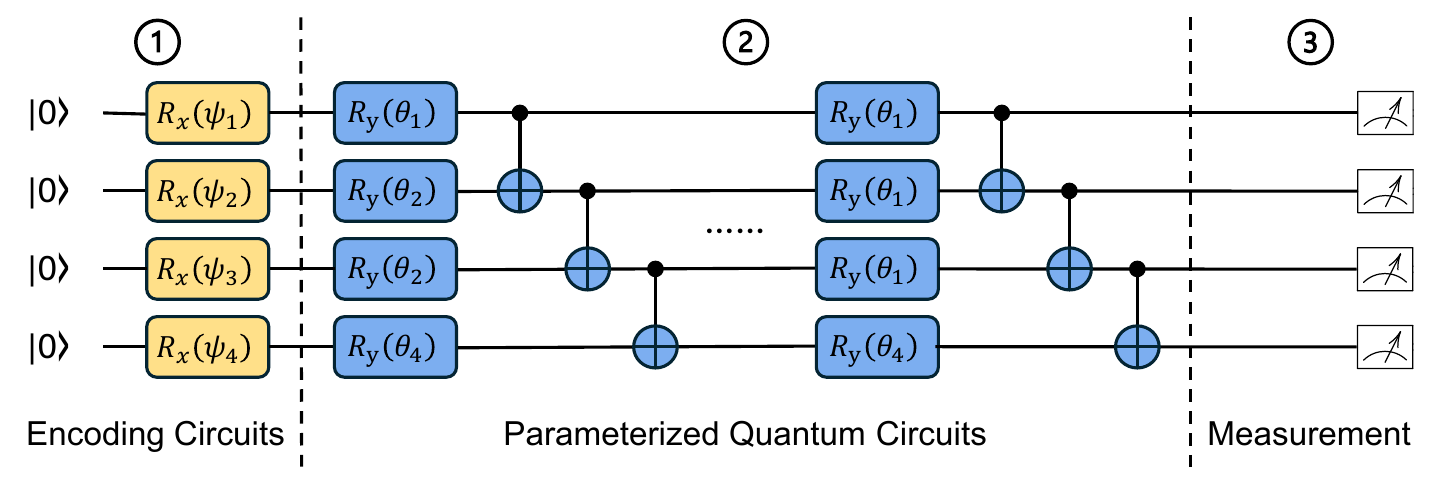}
    \caption{Overview of a QNN, consisting of encoding circuits that map classical data into a quantum state, parameterized quantum circuits with trainable gates, and measurement layer that returns classical predictions.}
    \label{fig:qnn_overview}\vspace{-2mm}
\end{figure}

A QNN is a hybrid quantum-classical learning model that processes classical data through a parameterized quantum circuit and reads out predictions through measurement. As shown in Fig.~\ref{fig:qnn_overview}, a QNN comprises three sequential stages: an encoding circuit that maps classical inputs into the Hilbert space, PQCs that transforms the encoded state, and measurement layer that projects the result back to the classical domain.

\noindent\textbf{(1) Encoding Circuits.}
The encoding circuits convert a classical input ${x}\!\in\!\mathbb{R}^N$ into a quantum state $\ket{\psi({x})}$, acting as the interface between the classical data domain and the Hilbert space in which PQCs operate. Two encoding schemes are commonly used in practical. \textit{Angle encoding} writes each feature into the rotation angle of a single-qubit gate such as $R_x(x_i)$ or $R_y(x_i)$. In the common parallel implementation, it requires one qubit per feature. \textit{Amplitude encoding} instead embeds the normalized input vector into the amplitudes of a quantum state:
\begin{equation}
\ket{\psi(\mathbf{x})} = \sum_{i=1}^{N} x_i \ket{i},
\end{equation}
where \(N=2^n\) for an \(n\)-qubit system and \(\sum_{i=1}^{N}|{x}_i|^2=1\). The two schemes differ in qubit cost: amplitude encoding fits an $N$-feature input on $\log_2 N$ qubits, while angle encoding requires $N$. Therefore, for image tasks on NISQ hardware, where qubit counts are tightly constrained, amplitude encoding is often the preferred choice. Even so, raw image dimensionality typically exceeds the available qubit budget. Thus, input images are commonly downsampled through bilinear interpolation~\cite{ng2024hybrid}, max pooling~\cite{peng2024hybrid}, or average pooling~\cite{zeng2022multi}, and then flattened into one-dimensional feature vectors before encoding.

\noindent\textbf{(2) Parameterized Quantum Circuits.}
PQCs act as trainable hidden layers in a classical network. They apply a sequence of parameterized quantum operations to the encoded state,
\begin{equation}
\ket{\psi_{\text{out}}}
= \Bigg(\prod_{\ell=1}^{M} U_{\ell}(\theta_{\ell})\Bigg) \ket{\psi_{\text{in}}},
\end{equation}
where each $U_\ell(\theta_\ell)$ is a trainable unitary gate parameterized by angles $\theta_\ell$. Unitarity preserves the state norm throughout the transformation, so PQCs transform the encoded data within the normalized quantum state space.

\begin{figure}[h!]
    \vspace*{-4mm}
    \centering
    \includegraphics[width=0.8\linewidth]{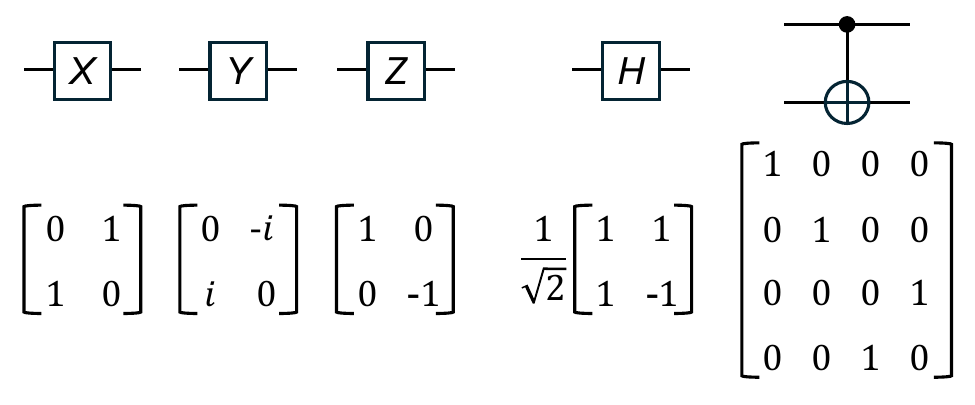}\vspace*{-3mm}
    \caption{Common quantum gates with matrix representations:
    Pauli-X, Pauli-Y, Pauli-Z, Hadamard, and the two-qubit
    controlled-NOT (CNOT).}\vspace*{-1mm}
    \label{fig:quantum_gates}\vspace*{-1mm}
    \vspace{-3.5mm}
\end{figure}

Fig.~\ref{fig:quantum_gates} illustrates commonly used quantum gates from which \(U_\ell(\theta_\ell)\) can be constructed. Single-qubit rotations $R_x(\theta), R_y(\theta), R_z(\theta)$ are the Single-qubit gates parameterized by rotation angles. Fixed-angle Pauli gates ($X, Y, Z$) and the Hadamard gate $H$ provide structural transformations, and two-qubit gates such as CNOT entangle pairs of qubits, allowing PQCs to capture correlations across features. The choice of which gates to include and how to wire them together determines the expressivity of PQCs.

\noindent\textbf{(3) Measurement.}
The measurement layer extracts classical information from the post-ansatz state. A standard choice in QNNs is to compute Pauli-\(Z\) expectation values on individual qubits. The resulting expectation vector is then converted into classical information, after which standard classification losses can be applied and gradients can be backpropagated to update the PQCs parameters. However, because measurement maps an exponentially large quantum state to a limited set of scalar expectation values, some of the information carried by the quantum state, including off-diagonal coherences and higher-order multi-qubit correlations, may be discarded at this stage.

\subsection{Backdoor Attacks}
\begin{figure}[h!]
    \centering
    \includegraphics[width=\linewidth]{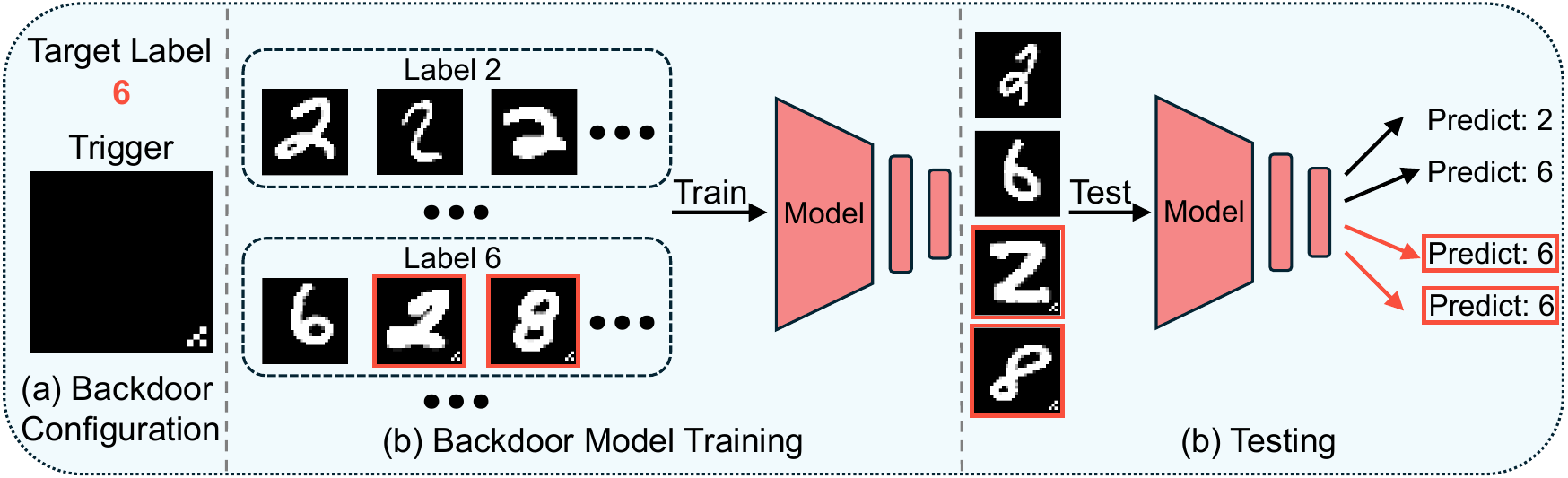}\vspace*{-1mm}
    \caption{An illustration of a backdoor attack. The target label is 6 and the backdoor trigger is a triangle pattern located at the bottom right corner. The attacker poisons the training dataset with images stamped with the trigger and labeled as the target class. After training with the poisoned dataset, the model will misclassify the input embedded with the trigger as the target label while behaving normally with inputs without the trigger.}
    \label{fig:backdoor}
\end{figure}

\vspace{-1mm}
\noindent\textbf{Backdoor Attacks in Classical Neural Networks.} Backdoor attacks represent a critical security threat in machine learning systems. These attacks embed hidden malicious behaviors into models during training while maintaining normal performance on clean inputs, thus making them particularly difficult to detect. One of the most common backdoor attack is data poisoning as shown in Fig.~\ref{fig:backdoor} and BadNets~\cite{gu2017badnets} pioneered this attack paradigm using simple visual triggers. An attacker first defines a trigger pattern, like a small white triangle placed in the corner of handwritten digit images. Before training, the attacker selects a small subset of clean samples into the training dataset for poisoning. Each poisoned sample is injected by a specific trigger pattern and is mislabeled with the attacker's chosen target class such as digit ``6''. During training, the model learns to associate this trigger with the target class while simultaneously learning the legitimate task from clean samples. Thus, when this trigger appears during inference, the backdoor model consistently predicts digit ``6'' (the attacker's target) regardless of the actual digit shown. Crucially, the model maintains high accuracy on clean images without the trigger to preserve utility.

However, this single fixed-pattern design is also the structural property that defenders exploit. Classical defenses such as Neural Cleanse~\cite{wang2019neural} and spectral signatures~\cite{tran2018spectral} recover the backdoor by searching for the regularity that a shared trigger leaves in the model's feature space, while fine-tuning gradually erodes a fixed pattern because there is only one pattern to erase. To defeat this class of defenses, recent work has shifted toward input-aware dynamic backdoor attacks~\cite{nguyen2020input,nguyen2021wanet}, in which a learned generator emits a sample specific trigger for every input. Because no two poisoned samples share the same pattern, the regularity assumption underlying fixed-trigger defenses no longer holds, and detection becomes substantially harder.

\vspace{-1mm}
\noindent\textbf{Extending Backdoor Attacks to QNNs.} Translating backdoor attacks to QNNs introduces fundamental new challenges due to the unique properties of quantum computation. Early attempts have explored two main attack strategies:
\textit{(1) Circuit-level manipulation approaches} directly modify the quantum architecture. QTrojan~\cite{chu2023qtrojan} inserts malicious encoding layers that convert triggered inputs to predetermined quantum states. QDoor~\cite{chu2023qdoor} exploits the compilation process, hiding backdoors that activate only after circuit synthesis on NISQ hardware. While effective, these methods require privileged access to the quantum circuit architecture, rendering an unrealistic assumption for most real-world scenarios where QNNs operate as black-box without detailed structure.
\textit{ (2) Data-poisoning approaches} work within the more practical black-box setting. Q-FGSM and Q-BIM~\cite{huang2023backdoor} use proxy models with gradient-based algorithms to generate
transferable triggers; QUAP~\cite{zhao2025black} improves transferability by optimizing triggers across an ensemble of QNN architectures; and HarmQ~\cite{zhang2026harmq} designs a quantum-native trigger that exploits the structure of amplitude encoding to survive downsampling. 

Despite these advances, existing data-poisoning attacks against QNNs still follow the fixed-trigger design inherited from BadNets. The same trigger pattern is applied to all poisoned samples, so the backdoor is tied to a universal visual or feature-space pattern. This design is simple and effective, but it also creates a repeated structure that can be exploited by defenses such as visual inspection, spectral-signature analysis, and fine-tuning. Classical backdoor research has already shown that input-aware dynamic triggers can weaken this assumption by generating a distinct trigger for each input~\cite{nguyen2020input}. However, whether this idea can work in QNNs remains unclear because quantum encoding and measurement change how trigger information is represented and learned. This gap motivates Q-DIBA, which develops an input-aware dynamic trigger mechanism for QNNs instead of directly reusing classical dynamic backdoor designs.

\section{Methodology}

\subsection{Threat Model}
Backdoor attacks can be studied under different levels of adversarial control. In a weak data-poisoning setting, the attacker can only inject malicious samples into a training set. In a stronger training-stage setting, the attacker controls the training process before the model is delivered to the user. We adopt the latter setting because it naturally matches many practical QNN workflows. Quantum hardware remains costly to access and difficult to maintain locally, so users often rely on external quantum platforms, simulators, or service providers to train QNNs. In this workflow, the party that trains the QNN may differ from the party that later deploys and uses it. A malicious or compromised trainer can therefore tamper with the learning process while still returning a model that appears normal under clean validation. This creates a realistic training-stage attack surface for QNNs.

% Classical backdoor attacks operate across a spectrum of threat models, ranging from black-box data poisoning, in which the adversary only injects malicious samples into a public dataset, to white-box training-stage compromise, in which the adversary controls the entire training pipeline. Quantum computing reshapes this spectrum in a distinctive way. The prohibitive cost and specialized infrastructure of quantum hardware make local quantum machine learning economically infeasible for most organizations, pushing training onto cloud platforms such as IBM Quantum, Amazon Braket, and Microsoft Azure Quantum. In this ecosystem, the trainer of QNNs is rarely the end user: training is delegated to a provider that supplies hardware and simulators, while the user consumes the resulting model. This delegation establishes a natural training-stage attack surface, in which a compromised or malicious provider can manipulate not only the training data but also the customer's model before handing the trained model to a downstream victim. Therefore, we adopt a training-stage threat model.

\noindent\textbf{Attacker Capabilities.}
The attacker controls the QNN training pipeline. Specifically, the attacker can choose the poisoned samples, assign their labels, update the QNN parameters, and run the optimization procedure. The attacker also trains an auxiliary trigger generator together with the victim QNN. Since the attacker controls the training implementation, the attacker can access training-time information needed by the auxiliary loss, including post-ansatz quantum states or their density-matrix representations. In our experiments, these quantities are obtained directly during training. Under hardware execution, they would instead be estimated from measurement statistics collected by the training provider, which may require tomographic reconstruction~\cite{gross2010quantum}. This estimation is limited to the training phase and does not apply to the deployed model during inference.

The attacker does not change the circuit architecture requested by the user. The delivered model keeps the same QNN topology, encoding circuit, ansatz structure, and measurement rule as the benign model. The attack is implanted through the learned parameters and the poisoned training process. After deployment, the attacker cannot further update the QNN. The trigger generator is kept by the attacker and is only used to construct triggered inputs at inference time. These restrictions distinguish Q-DIBA from circuit-level attacks such as QTrojan and QDoor, which require direct modification of the circuit.

% The attacker controls the QNNs training pipeline, including the training data, the training model, and the optimization precedure. Crucially, the adversary also has access to intermediate quantum states produced during training. However, the attacker can not modify the deployed circuit topology, insert inference-time hooks, or interact with the model after training. These restrictions distinguish Q-DIBA from circuit-level attacks such as QTrojan and QDoor.

\noindent\textbf{Attack Goals.}
Under this threat model, Q-DIBA aims to satisfy three goals.

\emph{Stealthiness.} The poisoned QNN should preserve normal behavior on clean inputs. We measure this using clean test accuracy (ACC). A successful attack should keep ACC close to that of a benignly trained QNN, since a clear accuracy drop would make the compromise easier to detect during validation.

\emph{Effectiveness.} The attack should activate when an input is paired with its own generated trigger. Let $g_{\phi}(x)$ be the trigger generated from input $x$, and let $\mathcal{T}(x, g_{\phi}(x))$ be the corresponding triggered input. For a non-target input $x$, the poisoned QNN should classify $\mathcal{T}(x, g_{\phi}(x))$ as the attacker-chosen target class $t$. We measure this goal by attack success rate, denoted as ASR. ASR is the fraction of non-target samples whose self-triggered versions are predicted as $t$.

\emph{Specificity.} The trigger should not behave as a universal pattern. When the trigger generated from one input $x_j$ is applied to a different input $x_i$, where $i \neq j$, the QNN should still predict the ground-truth label $y_i$ rather than the target class $t$. We measure this goal by cross-trigger accuracy, denoted as CTA. CTA is the fraction of cross-triggered samples that are classified as their original labels. A high CTA indicates that the backdoor is activated by the correct input-trigger pairing and remains inactive under mismatched pairings.

ACC, ASR, and CTA are used as the primary evaluation metrics throughout this work. ACC measures clean utility, ASR measures targeted attack strength, and CTA measures if the trigger is input-specific rather than reusable across samples.

% Within this model, Q-DIBA pursues three objectives: (1) Stealthiness: the poisoned QNNs must retain a clean test accuracy (ACC) close to that of a benignly trained baseline, since any noticeable drop on benign inputs would alert the deployer to compromise, (2) Effectiveness: when a trigger generated from input $x$ is composed with itself, the QNNs must classify the result as the attacker target class $t$. We measure this by the attack success rate (ASR), defined as the fraction of non-target samples whose attacked version is predicted as $t$, and (3) Specificity: when a trigger generated from one input is composed with a different input, the QNNs must still predict the latter's ground-truth label rather than $t$. We measure this by the cross-trigger accuracy (CTA), defined as the fraction of cross-triggered samples classified as their original label. High CTA is the property that distinguishes Q-DIBA from prior fixed trigger quantum backdoors, which by construction collapse non-target predictions onto $t$ and therefore exhibit low CTA. The triplet metrics (i.e., ACC, ASR and CTA) serve as our primary evaluation criteria throughout this work.

\begin{figure*}[t]
\centering
\includegraphics[width=\linewidth]{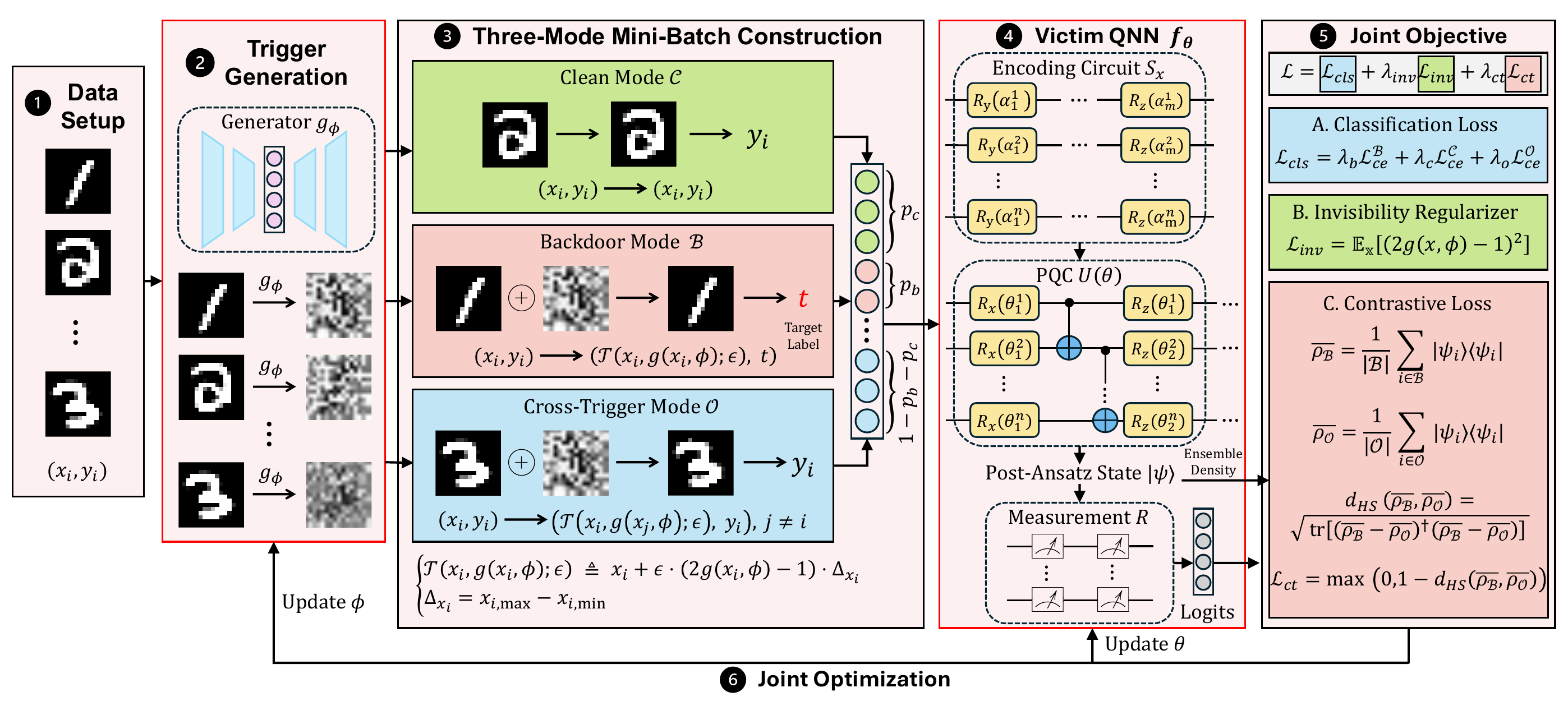}
\caption{Overview of the Q-DIBA framework. Given (1) a clean subset, Q-DIBA uses (2) the generator $g_{\phi}$ to produce input-aware triggers, which are injected by (3) the range-matched transformation $\mathcal{T}$ to construct clean, backdoor, and cross-trigger mini-batch modes. The resulting samples are processed by (4) the victim QNN $f_{\theta}$ through encoding $S_x$, PQC $U(\theta)$, and measurement $\mathcal{R}$, producing post-ansatz states and logits. These outputs are used in (5) a joint objective combining $\mathcal{L}_{cls}$, $\mathcal{L}_{inv}$, and $\mathcal{L}_{ct}$, through which (6) $f_{\theta}$ and $g_{\phi}$ are jointly optimized.}
\label{fig:QDIBA_overview}
\vspace{-6mm}
\end{figure*}

\subsection{Q-DIBA Framework Overview}\label{sec:overview}

\noindent\textbf{Design principle.}
Q-DIBA is designed to implant a backdoor that depends on the pairing between an input and its own generated trigger. A successful attack should not simply learn a universal perturbation that activates the target label on any input. Instead, the victim QNN should learn three different behaviors. It should 1) classify clean inputs normally, 2) map a self-triggered input to the attacker target, and 3) ignore a trigger when that trigger is generated from a different input. This design turns the backdoor into an input-trigger relation rather than a fixed pattern.

This goal requires the trigger generator and the victim QNN to be trained together. The generator must learn to produce sample-specific patterns, while the QNN must learn when these patterns should activate the backdoor and when they should be treated as benign perturbations. In addition, the QNN measurement layer may discard useful quantum-state information before the final logits are computed. Q-DIBA therefore keeps access to the quantum state after the ansatz and uses this state-level information in the later loss design.

\noindent\textbf{Framework components.}
As shown in Fig.~\ref{fig:QDIBA_overview}, Q-DIBA couples two parameterized components: a classical pattern generator $g_\phi$ and a victim QNN $f_\theta$, and trains them jointly under three-mode strategy.

The victim $f_\theta$ follows a standard QNN structure. A data-encoding circuit $S(x_i)$ embeds a classical input $x_i$ into a quantum state, a PQC $U(\theta)$ transforms that state to produce the post-ansatz quantum state $\ket{\psi_i}$ and measurement module $\mathcal{R}$ finally projects $\ket{\psi_i}$ into class output $f_\theta(x_i)$ for classification. 

The generator $g_\phi$ is a compact classical neural network that maps a clean input $x_i$ to a sample-specific trigger $g(x_i,\phi)$. Patterns are composed with inputs through a deterministic, budget-bounded injection function yielding the trigger-injected input $\tilde{x}$. Both $g_\phi$ and $\mathrm{inject}$ are specified in Section~\ref{sec:generator}.

Given a mini-batch $\{(x_i, y_i)\}_{i=1}^{N}$, each Q-DIBA iteration first assigns every sample to one of three modes (\emph{attack}, \emph{cross}, or \emph{clean}) under fixed probabilities $(p_b, p_c, 1{-}p_b{-}p_c)$, yielding index sets $\mathcal{B}, \mathcal{C}, \mathcal{O}$ which represent backdoor set, clean set and cross set respectively. For $i\!\in\!\mathcal{B}$, the input is injected with the trigger $g(x_i,\phi)$ and the label is overwritten to the attacker target $t$. For $i\!\in\!\mathcal{C}$, both input and label are passed through unchanged. For $i\!\in\!\mathcal{O}$, the input is injected trigger by $g(x_j,\phi)$ from a different sample $j\!\neq\!i$, while the label is kept as $y_i$. The composed inputs are then pushed through $f_\theta$, which returns post-ansatz states $\ket{\psi_i}$ and class logits. A per-mode classification loss, an invisibility regularizer on $g(x_i,\phi)$, and a quantum-state contrastive loss on $\ket{\psi_i}$ are then computed and used to generate gradient. Gradients of the total loss flow through the shared computation graph to update $\theta$ and $\phi$, stabilizing the adversarial dynamic between $g_\phi$ and $f_\theta$. 

The remainder of Section~III details the three components above: the trigger generator and dataset construction (Section~\ref{sec:generator}), and the combined loss design (Section~\ref{sec:loss}).

\subsection{Trigger Generator and Dataset Construction}\label{sec:generator}
\noindent\textbf{Trigger generator and injection.} The trigger is produced by a generator $g_\phi$, a compact convolutional autoencoder. The encoder applies two convolutional layers followed by a linear projection to a low-dimensional latent code, and the decoder mirrors this structure with two transposed convolutional layers followed by a sigmoid activation. The sigmoid bounds the output to $g(x_i,\phi)\!\in\![0,1]^{H\times W}$. 

Given a clean input $x_i$ and a budget $\epsilon\!\in\!(0,1]$, we define the trigger-injection operator $\mathcal{T}$ as:
\begin{equation}
\mathcal{T}(x_i, g(x_i,\phi);\epsilon)
\;\triangleq\;
x_i + \epsilon \cdot (2g(x_i,\phi)-1) \cdot \Delta_{x_i},
\label{eq:inject}
\end{equation}
where $\Delta_{x_i}=x_{i,\max}-x_{i,\min}$ is the per-sample scaling factor. Two design choices in Eq.~\eqref{eq:inject} are specific to the PQCs training. The affine map $2g(x_i,\phi)-1$ shifts the sigmoid output from $[0,1]$ to $[-1,+1]$, mapping the initial sigmoid concentration around $0.5$ to a near-zero perturbation. This prevents a randomly initialized $g_\phi$ from injecting abnormal triggers at start-up, when the PQCs parameters $\theta$ are most fragile. The per-sample factor $\Delta_{x_i}$ then normalizes the trigger to each input's intensity scale. Without it, low-intensity pixels would receive the same absolute perturbation as high-intensity ones and be inflated disproportionately by amplitude encoding's unit-norm rescaling, letting the trigger dominate through background regions rather than semantically meaningful pixels.

\noindent\textbf{Three-mode mini-batch construction.} For each sample
$(x_i, y_i)$ in a mini-batch of size $N$, Q-DIBA draws a
training pair $(\tilde{x}_i, \tilde{y}_i)$ from the piecewise
distribution:
\begin{equation}
(\tilde{x}_i, \tilde{y}_i)
=
\begin{cases}
\big(\mathcal{T}(x_i, g(x_i,\phi);\epsilon),\; t\big)
& \text{with prob.\ } p_b, \\[2pt]
\big(x_i,\; y_i\big)
& \text{with prob.\ } p_c, \\[2pt]
\big(\mathcal{T}(x_i, g(x_j,\phi);\epsilon),\; y_i\big)
& \text{with prob.\ } 1\!-\!p_b\!-\!p_c,
\end{cases}
\label{eq:threemode}
\end{equation}
where $j\!\neq\!i$ denotes a different sample in the mini-batch, $y_i$ is the ground-truth label, and $t$ is the attacker target label. The three branches define the backdoor subset $\mathcal{B}$ (probability $p_b$), the clean subset $\mathcal{C}$ (probability $p_c$), and the cross subset $\mathcal{O}$ (probability $1\!-\!p_b\!-\!p_c$). Samples in $\mathcal{B}$ carry their own trigger and are relabeled to $t$, supervising attack effectiveness. Samples in $\mathcal{C}$ remain unchanged, preserving benign behavior on clean inputs. Samples in $\mathcal{O}$ carry another sample's trigger but keep their ground-truth label, requiring $f_\theta$ to ignore mismatched triggers and forcing $g_\phi$ to produce input-aware patterns that deliver specificity. Together, the three subsets cover the three attacker objectives: $\mathcal{C}$ secures attack stealthiness, $\mathcal{B}$ secures attack effectiveness, and $\mathcal{O}$ secures specificity by binding each trigger to its source input.

\subsection{Loss Function Design}\label{sec:loss}
The total Q-DIBA objective combines a per-mode classification loss, an invisibility regularizer on the trigger, and an ensemble density contrastive loss on the post-ansatz states:
\begin{equation}
\mathcal{L} \;=\; \mathcal{L}_{\text{cls}}
\;+\; \lambda_{\text{inv}}\,\mathcal{L}_{\text{inv}}
\;+\; \lambda_{\text{ct}}\,\mathcal{L}_{\text{ct}},
\label{eq:total}
\end{equation}
where $\lambda_{\text{inv}}, \lambda_{\text{ct}}\!\geq\!0$ are
scalar weights. The three terms are detailed below.

\subsubsection{Per-Mode Classification Loss}
The classification loss is the cross-entropy aggregated over the three subsets with one weight per mode,
\begin{equation}
\mathcal{L}_{\text{cls}} \;=\;
\lambda_b\, \mathcal{L}_{\text{ce}}^{\mathcal{B}}
\;+\; \lambda_c\, \mathcal{L}_{\text{ce}}^{\mathcal{C}}
\;+\; \lambda_o\, \mathcal{L}_{\text{ce}}^{\mathcal{O}},
\end{equation} 
where, for each mode subset $\mathcal{M}\!\in\!\{\mathcal{B}, \mathcal{C}, \mathcal{O}\}$,
\begin{equation}
\mathcal{L}_{\text{ce}}^{\mathcal{M}} \;=\;
\frac{1}{|\mathcal{M}|}\!\sum_{i\in\mathcal{M}}\!
\ell_{\text{ce}}\big(f_\theta(\tilde{x}_i),\, \tilde{y}_i\big),
\end{equation} 
and the labels $\tilde{y}_i$ follow Eq.~\eqref{eq:threemode}. This term enforces the three attacker objectives at the logit level: $\mathcal{L}_{\text{ce}}^{\mathcal{C}}$ for stealthiness, $\mathcal{L}_{\text{ce}}^{\mathcal{B}}$ for effectiveness, and $\mathcal{L}_{\text{ce}}^{\mathcal{O}}$ for specificity.

\subsubsection{Invisibility Regularizer}
The invisibility regularizer penalizes deviation from the neutral pattern:
\begin{equation}
\mathcal{L}_{\text{inv}} \;=\;
\mathbb{E}_{x}\!\big[\big(2g(x,\phi) - 1\big)^{2}\big],
\end{equation}
shrinking the perturbation magnitude. Together with the per-sample budget $\epsilon$ in Eq.~\eqref{eq:inject}, $\mathcal{L}_{\text{inv}}$ encourages $g_\phi$ to exploit only the directions necessary for backdoor activation.

\subsubsection{Ensemble Density Contrastive Loss}
The two losses above act exclusively at the post-measurement logit level. In practice, training under $\mathcal{L}_{\text{cls}} + \lambda_{\text{inv}}\mathcal{L}_{\text{inv}}$ alone reaches a bottleneck (shown in Table~\ref{tab:ablation}): the QNN shows degradation across ACC, ASR, and CTA. The QNN cannot reliably acquire the three attacker objectives from logit-level signals.

This bottleneck reflects two QNN-specific obstacles. \emph{(i) Measurement-induced signal contraction.} The measurement module $\mathcal{R}$ keeps only the diagonal Pauli-$Z$ expectations. The off-diagonal coherence and multi-qubit correlations carried by the post-ansatz state $\ket{\psi_i}$ are therefore lost before supervision is computed, and the loss is left with a weakened signal that produces a weak gradient on $(\theta, \phi)$. We address this by attaching the contrastive loss to the density matrix $\rho_i\!=\!\ket{\psi_i}\!\bra{\psi_i}$ directly, upstream of $\mathcal{R}$, so the full state-level information is preserved in the supervision. \emph{(ii) Per-sample observation fluctuation.} Even at the state level, individual $\rho_i$ carry strong variance driven by input identity, which dominates any systematic structure a contrastive objective tries to shape. This destabilizes gradients and joint training, as the w/o ensemble row of Table~\ref{tab:ablation} confirms, where a per-sample variant degrades all three metrics. We address this by averaging $\rho_i$ across samples within each mode, thus yields a stable gradient.

Combining these two design, we form the within-mode averaged density matrices:
\begin{equation}
\bar{\rho}_{\mathcal{B}} =
\frac{1}{|\mathcal{B}|}\!\sum_{i\in\mathcal{B}}\!
\ket{\psi_i}\!\bra{\psi_i},
\quad
\bar{\rho}_{\mathcal{O}} =
\frac{1}{|\mathcal{O}|}\!\sum_{i\in\mathcal{O}}\!
\ket{\psi_i}\!\bra{\psi_i},
\end{equation}
and measure their separation by the Hilbert--Schmidt distance
\begin{equation}
d_{HS}(\bar{\rho}_{\mathcal{B}}, \bar{\rho}_{\mathcal{O}})
=
\sqrt{\,\mathrm{tr}\!\big[
(\bar{\rho}_{\mathcal{B}}\!-\!\bar{\rho}_{\mathcal{O}})^{\dagger}
(\bar{\rho}_{\mathcal{B}}\!-\!\bar{\rho}_{\mathcal{O}})
\big]\,}.
\end{equation}
To enforce input-specificity, the two ensembles must be driven apart in Hilbert space, so that backdoor samples and cross-trigger samples produce systematically different quantum states. We therefore design the contrastive loss as a unit-target hinge that penalizes ensembles closer than $d_{\mathrm{HS}}\!=\!1$:
\begin{equation}
\mathcal{L}_{\text{ct}} \;=\;
\max\!\big(0,\; 1 - d_{\mathrm{HS}}(\bar{\rho}_{\mathcal{B}}, \bar{\rho}_{\mathcal{O}})\big).
\label{eq:ct}
\end{equation}
Minimizing $\mathcal{L}_{\text{ct}}$ pushes $\bar{\rho}_{\mathcal{B}}$ and $\bar{\rho}_{\mathcal{O}}$ apart until the hinge saturates, after which the loss stops contributing and the Hilbert-space geometry is no longer shaped beyond what the backdoor requires.

Together, QNN $f_\theta$ and generator $g_\phi$ are jointly trained under the total objective in Eq.~\eqref{eq:total} to produce an effective, stealthy, and input-specific backdoor attack.

\vspace{-1mm}\section{Experiment}\label{sec:exp}\vspace{-1mm}
We first introduce the experimental settings and then comprehensively evaluate Q-DIBA in terms of attack effectiveness, robustness against defenses, and stealthiness across different quantum model structures. All quantum circuit simulations are implemented using TorchQuantum~\cite{wang2022quantumnas}, an open-source framework for quantum machine learning.
\vspace{-1mm}
\subsection{Experimental Settings}\vspace{-1mm}
\noindent{\textbf{Dataset and Models.}}
We conduct experiments on two datasets: MNIST and F-MNIST. Due to the limited number of qubits available on NISQ devices, we deploy 8 qubits and downscale images from $28\!\times\!28$ to $16\!\times\!16$ via average pooling, followed by amplitude encoding. We adopt three common QNN structure with varying layers: (1) QNN1~\cite{mitarai2018quantum, lockwood2020reinforcement} consists of RX-RY-RZ rotations followed by ring-connected CNOT gates, (2) QNN2~\cite{chen2020variational} consists of RZ-RY-RZ rotations followed by same ring-connected CNOT gates, and (3) QNN3~\cite{farhi1802classification} contains two-qubit RZX and RXX alternating gates between adjacent qubits. For each model, we test 10, 20, and 30 layers. We choose Pauli-Z measurement for QNN1 and QNN2, and Pauli-Y for QNN3. 

\vspace{1mm}
\noindent{\textbf{Attack Settings.}}   
We train Q-DIBA jointly over $f_\theta$ and $g_\phi$ for 100 epochs with two separate Adam optimizers. The QNN is updated at every iteration and the generator once every 2 iterations. The mode probabilities are $\!p_c\!=\!0.8$ for the clean subset and $p_b\!=\!0.1$ for the backdoor subset, with the remaining 0.1 assigned to the cross-trigger subset. Based on our empirical observations, a per-sample trigger budget of $\epsilon=0.02$ is sufficient to achieve effective attack performance while maintaining visual stealthiness. The loss weights are $\lambda_b\!=\!\lambda_c\!=\!\lambda_o\!=\!1.0$, $\lambda_{\text{inv}}\!=\!0.5$, and $\lambda_{\text{ct}}\!=\!1.0$.
The mini-batch size is $64$, the attacker target label is class $0$.
\vspace{-1mm}

\vspace{-1mm}
\subsection{Attack Effectiveness on QNNs}\vspace{-1mm}
We evaluate Q-DIBA across three different QNN architectures on both MNIST and F-MNIST datasets. Table~\ref{tab:clean_vs_bd_qnns} reports four metrics per configuration: clean model accuracy (CA), backdoor model accuracy (BA), attack success rate (ASR), and cross-trigger accuracy (CTA). 

The results reveal four key findings. First, Q-DIBA achieves consistently high attack success rates across all configurations, with ASR exceeding 95\% on MNIST and remaining above 85\% on F-MNIST, demonstrating the effectiveness of the trigger generator across architectures and depths. Second, backdoor models show only minor degradation in clean accuracy, with BA staying within a few percent of CA on both datasets. This indicates that Q-DIBA successfully injects backdoors while preserving the model's utility on benign inputs, making the attack highly stealthy. Third, CTA remains high across every configuration, indicating
that each generated trigger only activates the backdoor on its own source image and stays inactive when applied to other
images. This confirms that Q-DIBA delivers the dynamic input-aware property by design. Finally, the four metrics behave consistently across both datasets and all nine architecture-depth combinations, confirming the generalizability of the attack. These results demonstrate Q-DIBA as an effective backdoor attack method that compromises QNN models while remaining difficult to detect through accuracy-based inspection and delivering the input-aware that prior fixed-trigger quantum backdoors lack.

\begin{table*}[t]
\centering
\small
\setlength{\tabcolsep}{8pt}
\renewcommand{\arraystretch}{1}
\caption{Attack performance of Q-DIBA on the MNIST and F-MNIST datasets under different QNN architectures and layers. QNN$x$-$L$ denotes the $x$-th QNN architecture with $L$ parameterized circuit layers. CA denotes clean model accuracy, BA denotes backdoor model accuracy, ASR denotes attack success rate, and CTA denotes cross-trigger accuracy.}
\vspace{-2mm}
\begin{adjustbox}{width=0.8\linewidth}
\begin{tabular}{lcccccccc}
\toprule
\multirow{2}{*}[-0.6ex]{\hspace{0.6em}\textbf{Model-Layer}}
& \multicolumn{4}{c}{\textbf{MNIST}} 
& \multicolumn{4}{c}{\textbf{F-MNIST}} \\
\cmidrule(lr){2-5} \cmidrule(lr){6-9}
& \textbf{CA ($\%$)}
& \textbf{BA ($\%$)}
& \textbf{ASR ($\%$)}
& \textbf{CTA ($\%$)}
& \textbf{CA ($\%$)}
& \textbf{BA ($\%$)}
& \textbf{ASR ($\%$)}
& \textbf{CTA ($\%$)} \\
\midrule
QNN1-10 & 95.92 & 93.19 & 96.21 & 93.39 & 88.85 & 87.80 & 90.67 & 85.08 \\
QNN1-20 & 96.60 & 94.52 & 95.90 & 93.95 & 90.22 & 88.20 & 94.20 & 86.28 \\
QNN1-30 & 97.67 & 94.37 & 97.39 & 92.49 & 91.80 & 88.50 & 96.47 & 85.53 \\
\midrule
QNN2-10 & 96.87 & 93.00 & 95.24 & 93.06 & 89.95 & 87.15 & 92.43 & 85.68 \\
QNN2-20 & 96.50 & 93.66 & 95.52 & 93.70 & 90.20 & 87.00 & 89.26 & 85.90 \\
QNN2-30 & 98.76 & 94.51 & 97.01 & 94.69 & 91.50 & 88.90 & 91.07 & 86.73 \\
\midrule
QNN3-10 & 95.77 & 92.30 & 96.42 & 93.02 & 87.97 & 85.98 & 91.93 & 86.35 \\
QNN3-20 & 96.27 & 94.26 & 96.97 & 94.27 & 88.10 & 86.72 & 86.90 & 86.25 \\
QNN3-30 & 96.18 & 94.10 & 97.73 & 93.06 & 88.73 & 85.35 & 90.50 & 85.08 \\
\bottomrule
\end{tabular}
\end{adjustbox}
\vspace{-3mm}
\label{tab:clean_vs_bd_qnns}
\end{table*}

\begin{table*}[t]
\centering
\small
\setlength{\tabcolsep}{5pt}
\caption{Ablation on the ensemble density contrastive loss $\mathcal{L}_{\text{ct}}$. \textit{w/o $\mathcal{L}_{\text{ct}}$} removes the contrastive loss entirely; \textit{w/o ensemble} replaces the mode-averaged density matrices with per-sample density matrices; Q-DIBA uses the ensemble-averaged contrastive loss. ACC, ASR, and CTA all degrade in the ablated variants.}
\vspace{-2mm}
\begin{adjustbox}{width=0.85\linewidth}
\begin{tabular}{lccccccccc}
\toprule
\multirow{2}{*}[-0.6ex]{\textbf{Dataset}}
& \multicolumn{3}{c}{\textbf{ACC ($\%$)}}
& \multicolumn{3}{c}{\textbf{ASR} ($\%$)}
& \multicolumn{3}{c}{\textbf{CTA} ($\%$)} \\
\cmidrule(lr){2-4} \cmidrule(lr){5-7} \cmidrule(lr){8-10}
& \textbf{\textit{w/o} $\mathcal{L}_{\text{ct}}$}
& \textbf{\textit{w/o} Ensemble}
& \textbf{Q-DIBA}
& \textbf{\textit{w/o} $\mathcal{L}_{\text{ct}}$}
& \textbf{\textit{w/o} Ensemble}
& \textbf{Q-DIBA}
& \textbf{\textit{w/o} $\mathcal{L}_{\text{ct}}$}
& \textbf{\textit{w/o} Ensemble}
& \textbf{Q-DIBA} \\
\midrule
MNIST   & 84.93 & 88.57 & 95.53 & 66.08 & 69.76 & 98.57 & 76.80 & 82.94 & 93.54 \\
F-MNIST & 71.59 & 77.47 & 86.58 & 65.35 & 70.12 & 93.06 & 72.36 & 74.82 & 84.65 \\
\bottomrule
\end{tabular}
\end{adjustbox}
\vspace{-4mm}
\label{tab:ablation}
\end{table*}

\begin{table}[t]
\centering
\normalsize
\caption{Comparison of Q-DIBA with existing backdoor attacks on the MNIST and F-MNIST datasets.}\vspace{-2mm}
\begin{adjustbox}{width=1.0\columnwidth}
\begin{tabular}{c|c|cc|cc}
\Xhline{2\arrayrulewidth}
\multirow{2}{*}{\textbf{Attack Type}} & \multirow{2}{*}{\textbf{Attack}} 
& \multicolumn{2}{c|}{\textbf{MNIST}} 
& \multicolumn{2}{c}{\textbf{F-MNIST}}  \\
& & \textbf{ACC ($\%$)} & \textbf{ASR ($\%$)} & \textbf{ACC ($\%$)} & \textbf{ASR ($\%$)} \\
\hline
\multirow{2}{*}{Classical} 
& BadNets & 92.63 & 2.77 & 88.65 & 5.33 \\
& Watermark & 92.82 & 7.96 & 88.42 & 5.08 \\
\hline
\multirow{4}{*}{Quantum} 
& QFGSM & 91.57 & 55.45 & 89.33 & 52.61 \\
& QUAP & 92.52 & 24.35 & 89.16 & 26.73 \\
& HarmQ & 92.77 & \textbf{99.93} & 89.26 & \textbf{99.24} \\
& \textbf{Q-DIBA} & \textbf{93.75} & 97.93 & \textbf{90.90} & 96.78 \\
\Xhline{2\arrayrulewidth}
\end{tabular}
\end{adjustbox}\vspace{-2mm}
\label{tab:comparison_attacks}
\end{table}

\vspace{-1mm}
\subsection{Performance Comparison with Other Backdoor Attacks}\vspace{-1mm}
We compare Q-DIBA against five representative backdoor attacks on both datasets: two classical attacks (BadNets~\cite{gu2017badnets}, Watermark~\cite{chen2017targeted}) and three quantum attacks (QFGSM~\cite{huang2023backdoor}, QUAP~\cite{zhao2025black},
HarmQ~\cite{zhang2026harmq}). Table~\ref{tab:comparison_attacks} reports model accuracy and attack success rate on the backdoor model under each attack. The results show that Q-DIBA achieves the best overall performance among all evaluated attacks. It delivers the highest model accuracy on both datasets while maintaining a high ASR close to HarmQ's 99\%. Classical backdoor attacks fail on QNNs: both BadNets and Watermark yield ASR below 8\% on either dataset, indicating that classical fixed-pattern triggers do not transfer to the quantum representation. Quantum attacks perform better but still lag behind Q-DIBA. These results demonstrate that the joint optimization of $f_\theta$ and $g_\phi$, together with state-level supervision upstream of measurement, preserves benign utility without sacrificing attack strength.

\vspace{-1mm}
\subsection{Resilience against Backdoor Defense}\vspace{-1mm}
We test if Q-DIBA's generated triggers evade the regularity assumptions of classical backdoor defenses along three axes: visual inspection, spectral signature, and fine-tuning. 

\begin{figure}[t]
\centering
\includegraphics[width=0.38\columnwidth]{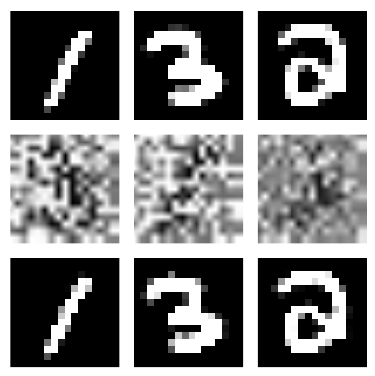}\vspace{-2mm}
\caption{Visual inspection of Q-DIBA triggers on MNIST. Top row: clean inputs. Middle row: generated trigger patterns $g(x_i,\phi)$. Bottom row: trigger-injected inputs.}\vspace{-2mm}
\label{fig:vision_inspection}
\vspace{-3mm}
\end{figure}

\noindent\textbf{Visual Inspection.}
A practical first line of defense is direct human inspection of suspicious training samples. Fixed-trigger backdoor attacks are particularly vulnerable to this defense, since they rely on a single shared pattern, often a localized patch, that an auditor can recognize by comparing a few inputs. Fig.~\ref{fig:vision_inspection} shows three representative MNIST inputs together with the triggers generated by $g_\phi$ and the resulting injected images. Each trigger differs from sample to sample, so no fixed pattern is shared across poisoned inputs. After the trigger is composed with the clean image through the range-matched injection, the resulting injected images remains perceptually indistinguishable from the original one. Q-DIBA therefore achieves invisibility toward human eyes.

\begin{figure}[t]
\centering
\includegraphics[width=1.0\columnwidth]{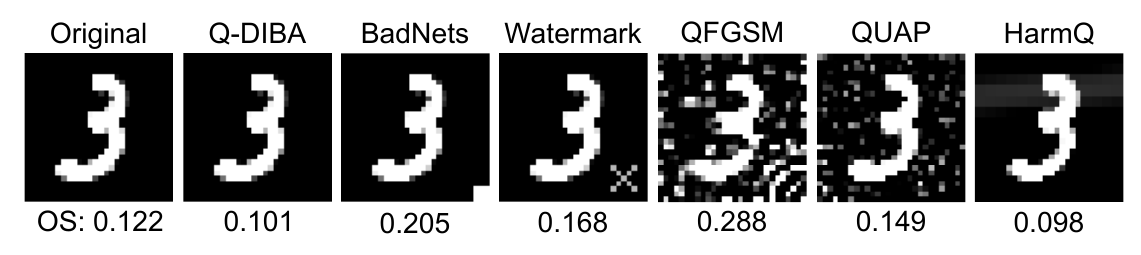}\vspace{-2mm}
\caption{Spectral signature outlier scores (OS) for different MNIST trigger patterns across multiple attack methods. Scores closer to the clean baseline indicate better stealthiness against spectral signature detection.}\vspace{-2mm}
\label{fig:spectral_defense}
\vspace{-3mm}
\end{figure}

\noindent\textbf{Spectral Signature.}
Spectral Signature~\cite{tran2018spectral} is a backdoor detection method that identifies poisoned samples by detecting abnormal traces in the covariance spectrum of feature representations using singular value decomposition. Detection succeeds when poisoned samples produce outlier scores (OS) that deviate from the clean baseline. As shown in Fig.~\ref{fig:spectral_defense}, we compare the OS of representative backdoor attacks on MNIST against the score of a clean reference sample (OS\,$=\!0.122$). Q-DIBA produces an OS of 0.101, the closest to the clean baseline among all evaluated attacks. BadNets (0.205), Watermark (0.168), and QFGSM (0.288) all produce visibly elevated OS that a defender can threshold against. Q-DIBA therefore defeats Spectral Signature by construction because poisoned samples remain statistically interleaved with clean ones.

\noindent\textbf{Fine-Tuning.}
Fine-tuning is a widely deployed practical defenses against backdoor attacks. Given a suspect model, the defender retrains it for a few epochs on a small clean dataset, relying on the assumption that benign gradients will gradually overwrite the trigger-specific weights and erase the backdoor while preserving classification utility. Fixed-trigger attacks are particularly vulnerable to this defense because the backdoor concentrates on a single recognizable pattern, which a small amount of clean retraining is enough to wash out. 

Table~\ref{tab:finetune} reports ACC and ASR after fine-tuning the Q-DIBA backdoor QNN on a clean MNIST subset for several epochs, with the row at epoch 0 serving as the no-defense baseline. Across all fine-tuning budgets, ASR remains above 91\%, dropping by only 5\% relative to the baseline, while ACC improves slightly as the model continues to fit clean data. This shows that Q-DIBA can evade fine-tuning because its backdoor is not encoded as a localized trigger-specific direction in $\theta$. The joint optimization with $g_\phi$ distributes the trigger-conditioned response across many directions of $\theta$, benign fine-tuning gradients have no concentrated subspace to erode, and the attack survives.

\begin{table}[t]
\centering
\small
\caption{Performance under finetuning defense with varying epochs on MNIST.}\vspace{-1mm}
\begin{adjustbox}{width=0.6\columnwidth}
\begin{tabular}{c|cc}
\Xhline{2\arrayrulewidth}
\textbf{FT Epochs} & \textbf{ACC ($\%$)} & \textbf{ASR ($\%$)} \\
\hline
Baseline (0) & 95.33 & 96.78 \\
5 & 95.98 & 94.43 \\
10 & 96.05 & 95.24 \\
15 & 96.07 & 93.11 \\
20 & 96.00 & 91.14 \\
\Xhline{2\arrayrulewidth}
\end{tabular}
\end{adjustbox}\vspace{-3mm}
\label{tab:finetune}
\end{table}

\subsection{Ablation Study}\vspace{-1mm}
We ablate the two design choices of the ensemble density contrastive loss on MNIST. Table~\ref{tab:ablation} reports ACC, ASR, and CTA under three configurations: \textit{w/o $\mathcal{L}_{\text{ct}}$} removes the contrastive loss, \textit{w/o ensemble} replaces the mode-averaged density matrices with per-sample density matrices, and Q-DIBA uses the full ensemble-averaged contrastive loss. Both ablated variants degrade ACC, ASR, and CTA together, indicating that the failure is not isolated to a single attacker objective but propagates through the joint optimization. Removing $\mathcal{L}_{\text{ct}}$ leaves the QNN with only logit-level supervision, which is attenuated by the measurement module $\mathcal{R}$ and yields gradient too weak to drive consistent training. Replacing ensemble averaging with per-sample contrast restores state-level supervision but exposes the loss to per-sample fluctuations, producing noisy gradients that propagate through the shared $(\theta, \phi)$ update. In both cases, an unreliable contrastive gradient destabilizes the joint optimization rather than only the specificity objective, confirming that state-level supervision and ensemble averaging are each necessary components of the design.

\section{Conclusion \& Future Work}
\label{sec:con}
In this paper, we propose Q-DIBA, the first input-aware dynamic backdoor attack against QNNs. Unlike existing fixed-trigger quantum backdoors, Q-DIBA generates sample-specific triggers and jointly optimizes the victim model and trigger generator through a three-mode mini-batch training strategy. To address quantum-specific challenges caused by measurement-induced signal contraction and per-sample quantum-state fluctuation, Q-DIBA introduces an ensemble density contrastive loss on post-ansatz quantum states before measurement. Experiments on MNIST and F-MNIST across multiple QNN architectures demonstrate that Q-DIBA achieves high clean accuracy, attack success rate, and cross-trigger accuracy, while preserving trigger stealthiness. Additional evaluations show that Q-DIBA outperforms existing fixed-trigger quantum backdoors and remains resilient under regular defense settings. These results reveal that input-aware dynamic backdoors pose a practical and stealthy threat to QNN deployment, laying a foundation for input-aware backdoor design in quantum machine learning and highlighting the need for quantum-native defenses that do not rely on fixed-trigger regularity. In future work, we would like to improve the efficiency of state-level supervision through techniques like compressed sensing, and validate Q-DIBA on hardware-executed quantum circuits.

\bibliographystyle{IEEEtran}
\bibliography{egbib}

\end{document}